\documentclass[english,sort&compress,reprint,superscriptaddress]{revtex4-1}
\usepackage{lmodern}

\usepackage[LGR,T1]{fontenc}
\usepackage[utf8]{inputenc}
\usepackage{textcomp}
\usepackage{amstext}
\usepackage{amssymb}
\usepackage{graphicx}
\usepackage{subscript}

\makeatletter

\ProvideTextCommand{\~}{LGR}[1]{\char126#1}

\usepackage{hyperref}
\usepackage{amssymb,amsmath}

\makeatother

\usepackage{babel}
\begin{document}

\title{Lateral heterostructures of hexagonal boron nitride and graphene:
BCN alloy formation and microstructuring mechanism}

\author{Marin Petrović}
\email{mpetrovic@ifs.hr}

\affiliation{Faculty of Physics and CENIDE, University of Duisburg-Essen, Lotharstr. 1,
D-47057 Duisburg, Germany}

\affiliation{Center of Excellence for Advanced Materials and Sensing Devices,
Institute of Physics, Bijenička cesta 46, HR-10000 Zagreb, Croatia}

\author{Michael Horn-von Hoegen}

\affiliation{Faculty of Physics and CENIDE, University of Duisburg-Essen, Lotharstr. 1,
D-47057 Duisburg, Germany}

\author{Frank-J. Meyer zu Heringdorf}

\affiliation{Faculty of Physics and CENIDE, University of Duisburg-Essen, Lotharstr. 1,
D-47057 Duisburg, Germany}
\begin{abstract}
Integration of individual two-dimensional materials into heterostructures
is a crucial step which enables development of new and technologically
interesting functional systems of reduced dimensionality. Here, well-defined
lateral heterostructures of hexagonal boron nitride and graphene are
synthesized on Ir(111) by performing sequential chemical vapor deposition
from borazine and ethylene in ultra-high vacuum. Low-energy electron
microscopy (LEEM) and selected-area electron diffraction (μ-LEED)
show that the heterostructures do not consist only of hexagonal boron
nitride (an insulator) and graphene (a conductor), but that also a
2D alloy made up of B, C, and N atoms (a semiconductor) is formed.
Composition and spatial extension of the alloy can be tuned by controlling
the parameters of the synthesis. A new method for in situ fabrication
of micro and nanostructures based on decomposition of hexagonal boron
nitride is experimentally demonstrated and modeled analytically, which
establishes a new route for production of BCN and graphene elements
of various shapes. In this way, atomically-thin conducting and semiconducting
components can be fabricated, serving as a basis for manufacturing
more complex devices.
\end{abstract}

\maketitle

\section{Introduction}

Two-dimensional materials (2DMs) have invoked significant interest
in the scientific community due to their exciting novel properties
and possible applications. Despite extensive investigation, incorporation
of 2DMs into new, more complex structures remains a challenge. One
attractive route for 2DMs integration is fabrication of heterostructures,
where layers of different materials are stacked one on top of another
or side by side \cite{Novoselov2016}. In this way, constituent materials
are exploited for obtaining thin and flexible systems with new functionalities
and improved properties. For example, heterostructures of graphene
(Gr), hexagonal boron nitride (hBN) and transition-metal dichalcogenides
can be used for fabrication of advanced transistors, diodes and photovoltaic
devices \cite{Dean2010a,Ross2014,Britnell2013,Duan2014}.

In the case of lateral heterostructures, a lot of attention has been
given to joint monolayers of Gr and hBN \cite{Ci2010}, since these
two materials both have hexagonal lattices with similar lattice parameters,
but differ substantially in terms of their electronic structure. While
Gr is an excellent conductor with high carrier mobility \cite{Bolotin2008},
hBN is an insulator with an electronic gap of $\sim6$ eV \cite{Cassabois2015}.
These properties make hBN-Gr lateral heterostructures very appealing
for construction of atomically thin electronics. However, it is still
demanding to fabricate these heterostructures as desired. One direction
of their production is the use of focused ion beam and lithography
(in combination with material regrowth) \cite{Levendorf2012,Liu2013,Gong2014a},
which are widely used for post-growth tailoring of the 2DMs. These
techniques represent straightforward tools for fabrication of hBN-Gr
interfaces of various shapes, but they require ex situ manipulation
of the 2DMs, resulting in a decreased quality of the final structures.
Also, control over the interface on the atomic scale (e.g., number
of defects, edge type or material intermixing) is poor, which can
impair functionality of future devices.

A common way to tackle these issues is by employing chemical vapor
deposition (CVD), which often implies epitaxial growth. For example,
atomically sharp interfaces in hBN-Gr lateral heterostructures have
been achieved on Rh(111), Ru(0001), Ir(111), Ni(111), and Cu foils
\cite{Sutter2012,Gao2013,Liu2014d,Sutter2014,Drost2015,Gao2015}.
However, heterostructures synthesized in such way were limited in
terms of shape and size, and no real-time monitoring of their fabrication
was possible in most cases. An exception is hBN-Gr synthesis on Ru(0001),
where low-energy electron microscopy (LEEM) was used to uncover details
of growth dynamics \cite{Sutter2012,Sutter2014}. In a general case,
atomically sharp interfaces between hBN and Gr may not always be desirable
and a certain level of mixing (i.e. 2D alloying) can be useful. In
particular, a material consisting of hexagonally arranged C, B and
N atoms (hBCN, without implication of exact stoichiometry) was theoretically
predicted to be semiconducting with a bandgap tunable by altering
the C:(B,N) ratio \cite{Lam2011,DaRochaMartins2011,Manna2011,Xie2012,Zhang2015g}.
These predictions were experimentally confirmed for the case of hBC\textsubscript{2}N
where a direct bandgap of $\approx2$ eV was found \cite{Watanabe1996,Lu2013},
which makes this alloyed material very interesting for optoelectronic
applications. Therefore, hBCN holds a unique place in the family of
2D materials, and its investigation is well-justified.

Here, we study hBN-Gr lateral heterostructures on Ir(111). By using
LEEM, the synthesis was monitored in real-time, and details of the
hBN-Gr boundary were investigated by low-energy electron diffraction
(LEED). In addition, we introduce a technique which effectively produces
various micron-sized elements of graphene and hBCN alloy out of heterostructures,
establishing in such way a new route for microstructuring of 2DMs.

\section{Experimental Methods}

Growth of hBN-Gr heterostructures was conducted on a Ir(111) surface
(single-crystal, Mateck) in an ultra-high vacuum (UHV) setup dedicated
to LEEM and photoemission electron microscopy (PEEM) at the University
of Duisburg-Essen. The Ir crystal was cleaned by Ar\textsuperscript{+}
sputtering at 2 keV followed by several cycles of heating in oxygen
at 1170 K and annealing at 1470 K. Sequential CVDs from borazine (B\textsubscript{3}H\textsubscript{6}N\textsubscript{3})
and ethylene (C\textsubscript{2}H\textsubscript{4}) were used for
the synthesis of hBN-Gr heterostructures. Experimental parameters
used in the synthesis and microstructuring process are discussed in
the context of the specific experiments below. The sample temperature
was measured with an infrared pyrometer. Borazine and ethylene were
dosed into the UHV chamber through leak-valves. Borazine was kept
in a Peltier cooler to prevent its degradation. Prior to each hBN
synthesis, the borazine inlet line was evacuated to remove any decomposition
residues.

LEEM, LEED and PEEM measurements were carried out in situ with an
Elmitec SPE-LEEM III microscope with a spatial resolution of $\sim10$
nm. A mercury discharge lamp was used for sample illumination during
PEEM. The bright field (BF) mode was used for LEEM imaging, where
the contrast aperture allowed passage of the central (0,0) spot and
the first-order moiré spots into the imaging column of the microscope.
Consequently, areas on the sample exhibiting even the smallest differences
in the moiré pattern can be imaged with distinctive contrast in real
space, and this will be very useful for visualization of different
regions of hBN-Gr heterostructures. The same effect has been used
for identification of local lattice constant changes of Gr on Ir(111)
during wrinkling \cite{Petrovic2015}. In electron diffraction experiments,
only a small micron-sized circular area of the sample surface ($d\approx0.5$
μm) was illuminated with an electron beam, thus enabling micro-LEED
(μ-LEED) analysis.

\section{Results}

\subsection{Heterostructure Growth and Composition}

\begin{figure}
\begin{centering}
\includegraphics[width=8.4cm]{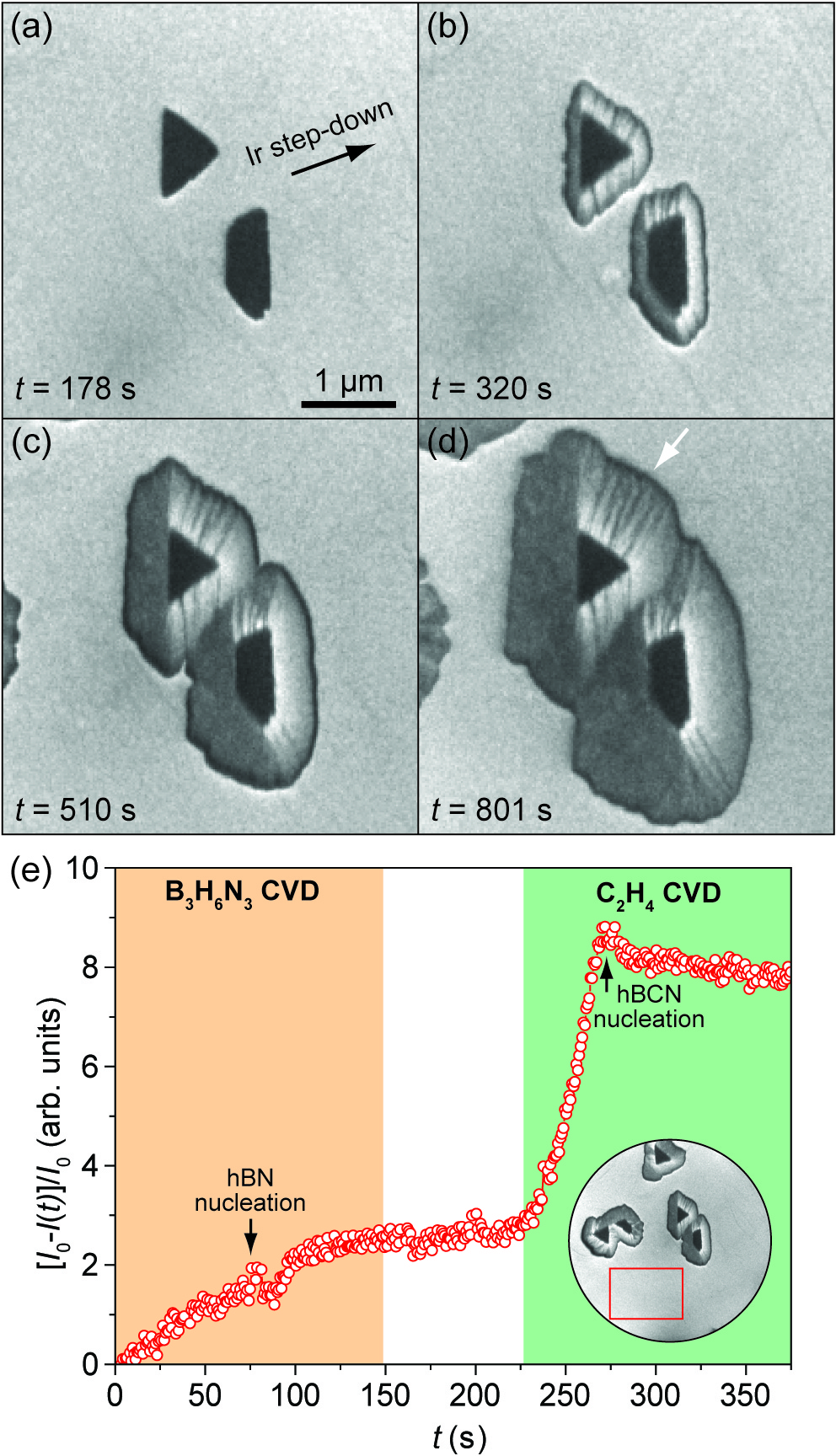}
\par\end{centering}
\caption{\label{fig1}(a)-(d) A sequence of LEEM images showing the formation
of hBN-Gr lateral heterostructures. In panel (a), two hBN islands
are visible. In panels (b)-(d), new material (Gr and hBCN alloy) is
growing from the edges of hBN islands. Time relative to the beginning
of the borazine dosing is indicated in each panel. $T_{\mathrm{S}}=1170$
K, $E=9$ eV. (d) Normalized change of electron reflectivity as a
function of time. Data was extracted from the area indicated by a
rectangle in the inset, in the vicinity of the heterostructures shown
in (a)-(d). Orange and green shading indicate periods of borazine
and ethylene CVD, respectively.}
\end{figure}

The formation of hBN-Gr lateral heterostructures is shown in Figure
1 in a sequence of LEEM images (see also Supplementary Movie 1). First,
well-defined hBN islands are grown as shown in Figure \ref{fig1}(a)
by exposing the Ir surface at 1170 K to $10^{-8}$ mbar of borazine.
As reported previously, triangular and trapezoidal hBN islands form
on the surface, with the short base of the trapezoidal islands oriented
in the step-up direction of the Ir surface \cite{Petrovic2017a}.
After stopping the borazine dosing and allowing the pressure to recover,
ethylene was introduced into the UHV chamber as a precursor for Gr
growth with a pressure of $10^{-8}$ mbar, while keeping the sample
temperature at 1170 K the entire time. This resulted in the formation
of the new material exclusively at the edges of preexisting hBN islands,
as shown in Figure \ref{fig1}(b), whose structure and composition
will be analyzed later in the text. If the ethylene pressure during
CVD was raised to $\sim10^{-7}$ mbar, additional Gr nucleation on
bare Ir areas was observed. Heterostructure growth in the vertical
direction is not possible due to the inertness (i.e., lack of catalytic
activity) of hBN. Further exposure to ethylene results in roughly
isotropic expansion of the heterostructure perimeter, as is visible
in Figure \ref{fig1}(c) and (d). However, the new material growing
attached to hBN exhibits variations in electron reflectivity, indicating
that the material is not purely Gr, as might be expected during dosage
of solely ethylene. PEEM images of heterostructures also show similar
variations of intensity, indicating alteration of the electronic properties
across the heterostructure (see Supplementary Figure S1). In all our
LEEM data, bright regions always form in the step-down direction of
the Ir surface. Also, streaks of dark regions, often confined by the
Ir step edges, are regularly found embedded in the bright regions
{[}see white arrow in Figure \ref{fig1}(d){]}. These findings indicate
that the Ir surface step morphology is decisive for the growth of
the new material.

The concentration of adsorbed species on the Ir surface during heterostructure
formation was determined by monitoring LEEM reflectivity in the vicinity
of the growing islands. The concentration of adsorbates $c_{\mathrm{ad}}(t)$
is proportional to the LEEM reflectivity decrease $I_{0}-I\left(t\right)$,
where $I_{0}$ is reflectivity of the surface free of any adsorbates
\cite{Loginova2008,Sutter2012}. Normalized reflectivity change $\left[I_{0}-I\left(t\right)\right]/I_{0}\propto c_{\mathrm{ad}}\left(t\right)$
is represented in Figure \ref{fig1}(e) by red circles, with the corresponding
data taken from the area indicated with a red rectangle in the inset.
The concentration of the adsorbates increases monotonically upon borazine
introduction into the UHV chamber until the borazine is turned off
{[}orange region in Figure \ref{fig1}(e){]}, with a small drop at
$t\approx75$ s which we relate to the onset of hBN nucleation \cite{Loginova2008}.
It is crucial to note that there are still adsorbate species present
on the Ir surface even after the borazine supply has been switched
off and before the ethylene is introduced into the chamber at $t\approx225$
s {[}green region in Figure \ref{fig1}(e){]}. The concentration of
all adsorbed species increases rapidly until $t\approx275$ s when
the formation of new material attached to hBN cores starts, which
is visible in Figure \ref{fig1}(e) as a sharp onset of the concentration
decrease. Because different adsorbates generally cause different reflectivity
changes in LEEM, comparison of adsorbate concentrations during the
course of heterostructure formation in Figure \ref{fig1}(e) cannot
be done directly.

\begin{figure}
\begin{centering}
\includegraphics{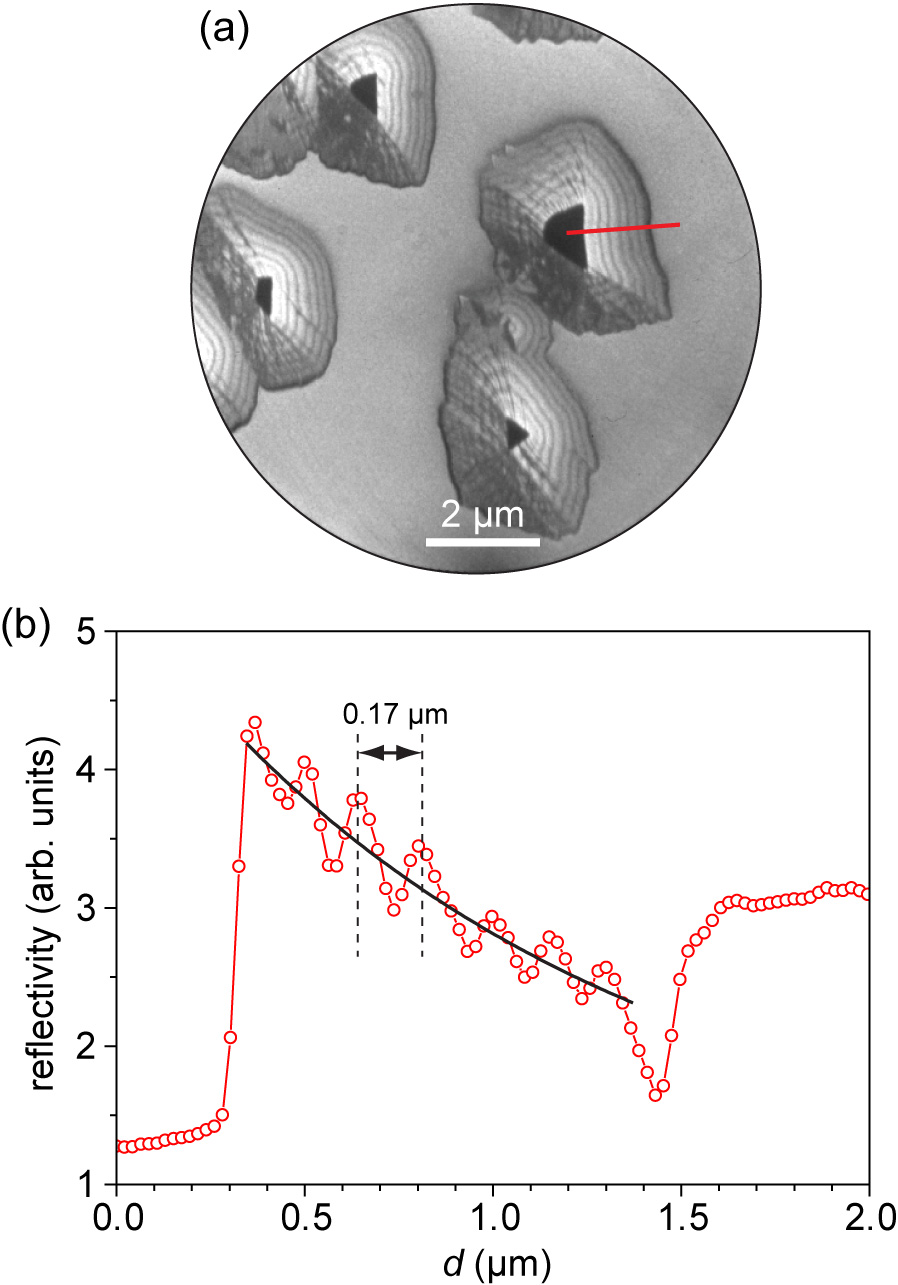}
\par\end{centering}
\caption{\label{fig2}(a) LEEM image of the hBN-Gr heterostructure with modulated
carbon content {[}i.e., C:(B,N) ratio{]} in the region of the hBCN
alloy, visible as equidistant contours (dark contours correspond to
carbon-rich regions, bright contours to carbon-poor regions). $T_{\mathrm{S}}=1170$
K, $E=9$ eV. (b) Cross-section taken from panel (a) as indicated
by a red line. Carbon-rich and carbon-poor regions are identified
as minima and maxima in LEEM reflectivity that are superposed to an
exponentially decaying background (black line). Red lines are guides
to the eye.}
\end{figure}

The reflectivity analysis suggests that during sequential CVD growth
of hBN-Gr heterostructures, mixing of species originating from both
borazine and ethylene takes place on the Ir surface and an alloy is
formed. Further clear-cut evidence of this is found in experiments
where the ethylene pressure was periodically modified during CVD between
$2\times10^{-8}$ and $7\times10^{-8}$ mbar with a period of $\sim42$
s, thus varying the carbon supply during the synthesis. The resulting
structure is shown in Figure \ref{fig2}(a), where hBN cores are surrounded
by equidistant carbon-poor (high reflectivity contours) and carbon-rich
(low reflectivity contours) regions. A cross-section taken from Figure
\ref{fig2}(a) is shown in panel (b), where six minima corresponding
to six consecutive ethylene pressure increases are superposed to exponentially
decaying LEEM reflectivity (see also Supplementary Figure S2). Lateral
extension of carbon-poor or carbon-rich regions can be tuned by changing
the duration of decreased or increased ethylene pressure in the course
of the synthesis. For the heterostructures shown in Figure \ref{fig2},
average separation of the individual regions is less than 0.2 μm.

If the sample is treated with oxygen between the borazine and ethylene
CVD steps, the concentration of the leftover borazine species is significantly
reduced. Consequently, pure hBN and Gr regions become better defined
and the transition region between them becomes narrower (see Supplementary
Figure S3). Another route for controlling the extent of the transition
region is modification of the synthesis temperature, where lower temperatures
produce narrower transition between hBN and Gr due to the lower concentration
of leftover adsorbed borazine species (see Supplementary Figure S2).
The same procedures for improving Gr-hBN interface are effective for
heterostructures synthesized on Ru(0001) \cite{Sutter2012}. At this
point we assume (and later on we will elucidate) that the mixed material
is a hexagonal mesh composed of C, B and N, and hence we label it
as hBCN.

\begin{figure*}
\begin{centering}
\includegraphics[width=15.4cm]{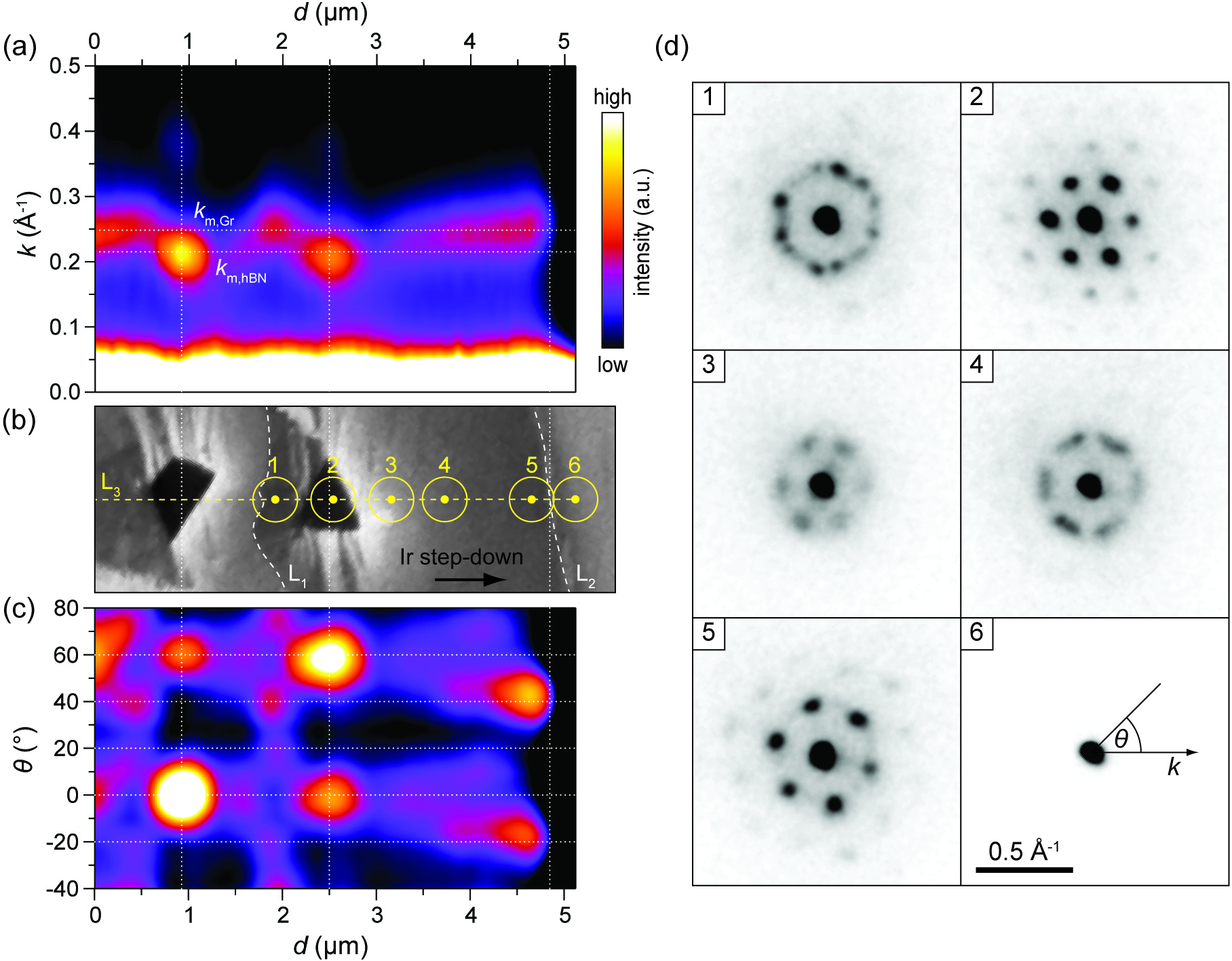}
\par\end{centering}
\caption{\label{fig3}Analysis of hBN-Gr heterostructures with μ-LEED. Analyzed
area of the sample is shown in panel (b). Line L\protect\textsubscript{1}
marks the border between the two coalesced heterostructures, line
L\protect\textsubscript{2} marks the edge of a heterostructure towards
Ir, and line L\protect\textsubscript{3} marks the path of the μ-LEED
scan. Yellow circles designate characteristic areas whose diffraction
patterns are shown in panel (d) {[}only vicinity of the $\left(0,0\right)$
spot is shown{]}. Panels (a) and (c) show spatially-dependent radial
average and polar profile of the of μ-LEED patterns. Spatial coordinates
of panels (a), (b) and (c) are aligned. $E=9.2$ eV in LEEM, $E=35.5$
eV in LEED.}
\end{figure*}

The crystal structure of the hBN-Gr heterostructures was analyzed
in detail by performing μ-LEED. The area of the sample selected for
this is shown in Figure \ref{fig3}(b) where two coalesced heterostructures
are visible, similar as in Figure \ref{fig1}(d). The boundary between
them is highlighted by the dashed line L\textsubscript{1}. The edge
towards uncovered Ir is marked by the dashed line L\textsubscript{2}.
The path of the μ-LEED scan is marked by the horizontal dashed line
L\textsubscript{3}. Important information about the crystal structure
was extracted from the analysis of the central region of the diffraction
pattern where the moiré diffraction spots surrounding the $\left(0,0\right)$
spot are located. Several characteristic diffraction patterns, originating
from the areas marked in Figure \ref{fig3}(b) by yellow circles,
are shown in Figure \ref{fig3}(d) (a complete μ-LEED scan is given
in the Supplementary Movie 2). Diffraction of the core hBN island
(box 2) shows a clear pattern corresponding to hBN/Ir(111). The diffraction
recorded at the edge of the heterostructure (box 5) exhibits the pattern
which is associated with Gr/Ir(111). Between these two limiting cases,
in the hBCN region the diffraction pattern shows significant variation
of the moiré spots position and intensity (boxes 1, 3 and 4). To capture
this variation in a systematic way, radial average and polar profile
of the central part of the μ-LEED scan were extracted from the Supplementary
Movie 2 and are plotted in Figure \ref{fig3}(a) and (c), respectively.
The radial average displays information about the distance of the
moiré diffraction spots from the center of the Brillouin zone (i.e.,
their \textit{k}-value), and the polar profile provides information
about the angular distribution of the moiré spots. Calibration of
the \textit{k}-space was done by measuring the first-order Ir(111)
diffraction spots ($k_{\mathrm{Ir}}$, $2\pi/2.715\:\mathrm{\mathring{A}}$
\cite{NDiaye2008}).

Small changes in the rotation or lattice size of epitaxial 2D materials
are effectively magnified by an order of magnitude via moiré spots
re-positioning \cite{NDiaye2008}. This effect is exploited in the
following, where we focus on the right one of the two hBN-Gr heterostructures
from Figure \ref{fig3}(b). Saturated intensity at the bottom of the
Figure \ref{fig3}(a) corresponds to the $\left(0,0\right)$ spot,
which is present throughout the μ-LEED scan (including the area of
bare Ir). The second most intense peak corresponds to the first-order
moiré diffraction spots. Evidently, the distance of the moiré peak
from the $\left(0,0\right)$ spot changes across the scan, as well
as its intensity. The \textit{k}-values of the Gr ($k_{\mathrm{m,Gr}}$,
$2\pi/25.3\:\mathrm{\mathring{A}}$) and hBN ($k_{\mathrm{m,hBN}}$,
$2\pi/29.1\:\mathrm{\mathring{A}}$) moiré structures on Ir(111) \cite{NDiaye2008,FarwickZumHagen2016}
are marked by horizontal dotted lines. The moiré peak shifts between
these two values as a function of position within the heterostructure.
Directly at the hBN core, the first-order moiré diffraction spots
have a \textit{k}-value corresponding to the hBN/Ir(111), in agreement
with a clear and sharp diffraction pattern shown in box 2 of Figure
\ref{fig3}(d). In a similar manner, material located at the edge
of the heterostructure is characterized by a \textit{k}-value of Gr/Ir(111)
and shows a diffraction pattern characteristic for that system {[}box
5 of Figure \ref{fig3}(d){]}. When scanning from the hBN to the Gr
region (in the step-down direction of the Ir surface), the \textit{k}-value
of the moiré structure gradually increases, i.e., the real space lattice
constant of hBCN decreases. Also, the intensity of the moiré diffraction
spots drops suddenly after leaving the hBN core {[}cf. box 3 of Figure
\ref{fig3}(d){]}, and gradually increases as the edge of the heterostructure
is reached. These changes reflect the quality of the moiré structure
in the examined areas in real space, i.e., moiré uniformity in terms
of periodicity and amplitude.

Valuable information about the local orientation of the grown heterostructure
is gained from polar profiles extracted from the μ-LEED scan. They
are shown in Figure \ref{fig3}(c) for an angular range of 120\textdegree{}
(the diffraction patterns are 3-fold symmetric \cite{Orlando2014,Petrovic2017a}),
where 0\textdegree{} corresponds to an alignment of the zig-zag direction
of the hBN and dense packed rows of Ir(111). Focusing again on the
right one of the two heterostructures in Figure \ref{fig3}(b), the
moiré diffraction pattern of the hBN core exhibits two distinctive
peaks, at 0\textdegree{} and 60\textdegree , where one peak is more
intense than the other one, due to the two different atoms in the
basis of the hBN lattice. When going in the step-down direction towards
the heterostructure edge, the crystal structure exhibits several rotations
as well as spot splitting, indicating that the transition from hBN
to Gr is nontrivial, i.e., it contains hexagonal lattices of different
orientation and composition. The maximum observed rotation of the
moiré pattern with respect to the crystallographic orientation of
the hBN lattice is $\sim20$\textdegree , which translates into real
rotations of $\lesssim2\text{\textdegree}$ when the magnifying property
of the moiré is taken into account. We note that due to the nature
of the moiré structure, its rotation and lattice parameter are interlinked.
In particular, 20\textdegree{} rotation could cause up to $\sim6$
\% lattice parameter increase, which is insufficient to account for
our observations (15\% lattice parameter increase) described in the
preceding paragraph. In the Ir step-up direction, the boundary between
hBN and Gr is sharper and does not show obvious signs of alloying
{[}see Figure \ref{fig1} or Figure \ref{fig3}(b){]}. However, a
detailed analysis presented in Figure \ref{fig3} points to the existence
of a narrow alloy region in the vicinity of the hBN-Gr boundary. The
moiré \textit{k}-value shifts from $k_{\mathrm{m,hBN}}$ to $k_{\mathrm{m,Gr}}$,
and also similar relative lattice rotations up to $\sim2\text{\textdegree}$
are found.

\subsection{Microstructuring via hBN Disintegration}

\begin{figure}
\begin{centering}
\includegraphics[width=8.4cm]{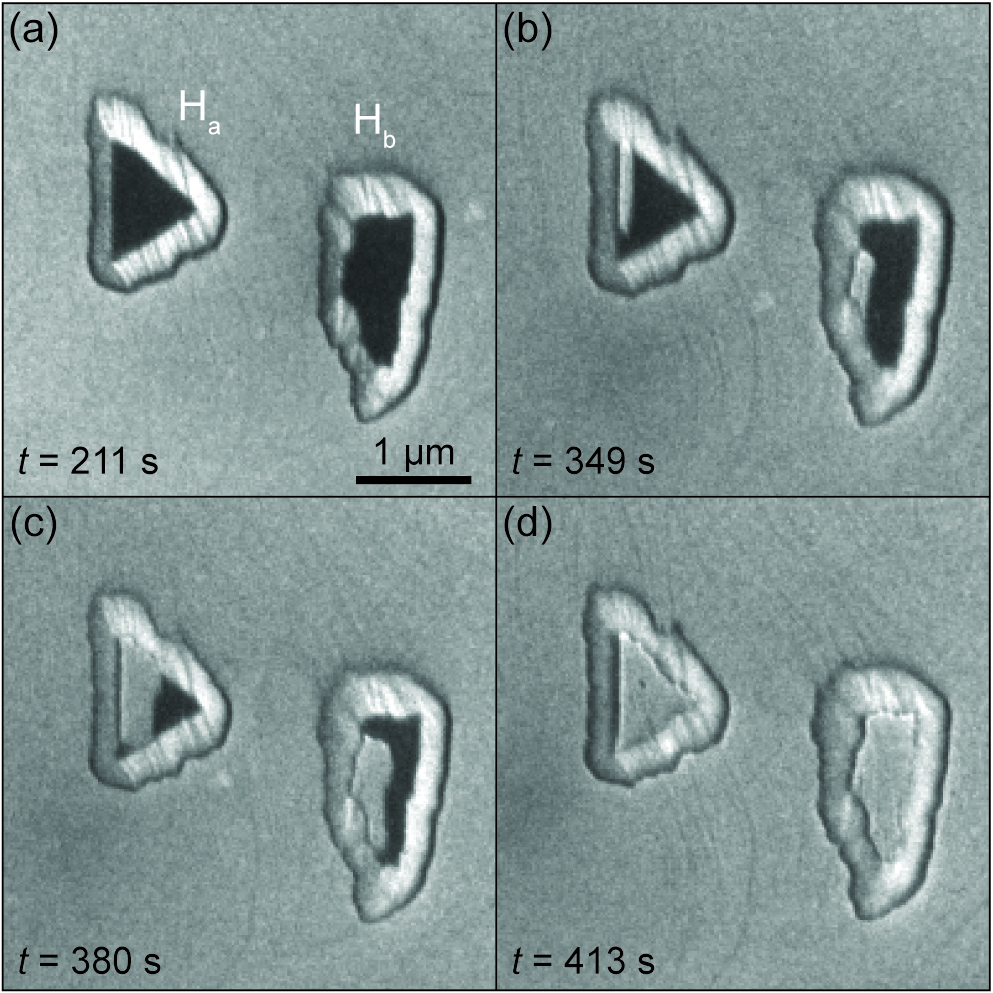}
\par\end{centering}
\caption{\label{fig4}Formation of hollow 2D microstructures via hBN disintegration
at high temperatures. (a) Well-defined hBN-Gr heterostructures H\protect\textsubscript{a}
and H\protect\textsubscript{b} are synthesized at 1170 K. (b)-(d)
Annealing of heterostructures at 1370 K results in disintegration
of hBN, while leaving the surrounding hBCN and Gr intact. Time relative
to the beginning of the borazine dosing is indicated in each panel.
$E=9$ eV.}
\end{figure}

If desired, hBN-Gr heterostructures with a narrow hBCN belt surrounding
an hBN core can be synthesized through a short ethylene CVD step,
in a similar way as has been done on other substrates \cite{Han2013,Sutter2014}.
An example is shown in Figure \ref{fig4}(a) for a synthesis at 1170
K where the borazine has been dosed at $10^{-8}$ mbar for 129 s and
ethylene at $2\times10^{-8}$ mbar for 58 s, resulting in two isolated
heterostructures H\textsubscript{a} and H\textsubscript{b}. Then
the heterostructures were subsequently heated up to 1370 K at a rate
of $\sim1.5$ K/s, and the temperature was kept at that value for
several minutes. During the entire heating and annealing process,
the heterostructures H\textsubscript{a} and H\textsubscript{b} were
being imaged with LEEM (see Supplementary Movie 3). Due to significant
decomposition at temperatures higher than $\sim1220$ K \cite{Usachov2012,FarwickZumHagen2016,Petrovic2017a},
hBN embedded within the heterostructures starts disintegrating as
shown in Figure \ref{fig4}(b). The dominant locations for initiation
of disintegration are boundaries between hBN and hBCN, which can be
linked to the fact that interaction of the heterostructure with the
Ir substrate is stronger there \cite{Liu2014d}. As the sample is
kept at high temperature, disintegration continues exclusively at
the free hBN edge that is not in contact with the rest of the heterostructure
{[}Figure \ref{fig4}(c){]}, and eventually results in a complete
removal of the hBN from the heterostructure {[}Figure \ref{fig4}(d){]}.
In this process, the surrounding material remains intact (at least
on the scales accessible to LEEM), and the final result are hollow
2D microstructures of hBCN and Gr. Additional evidence for complete
removal of hBN core is found in PEEM measurements, in which only the
remaining microstructures are visible due to the reduction of the
work function (see Supplementary Figure S1). Size and lateral ``thickness''
of these microstructures depend on the size of the initial hBN cores
and the lateral extension of the surrounding hBCN. Since hBN grows
on Ir either as triangular or trapezoidal islands, the shape of Gr
and hBCN microstructures is limited to a certain extent. However,
subsequent oxygen etching of microstructures as well as their additional
growth by ethylene CVD provides routes for further modification of
their shape and size (an example will be given further below).

\begin{figure*}
\begin{centering}
\includegraphics[width=13.6cm]{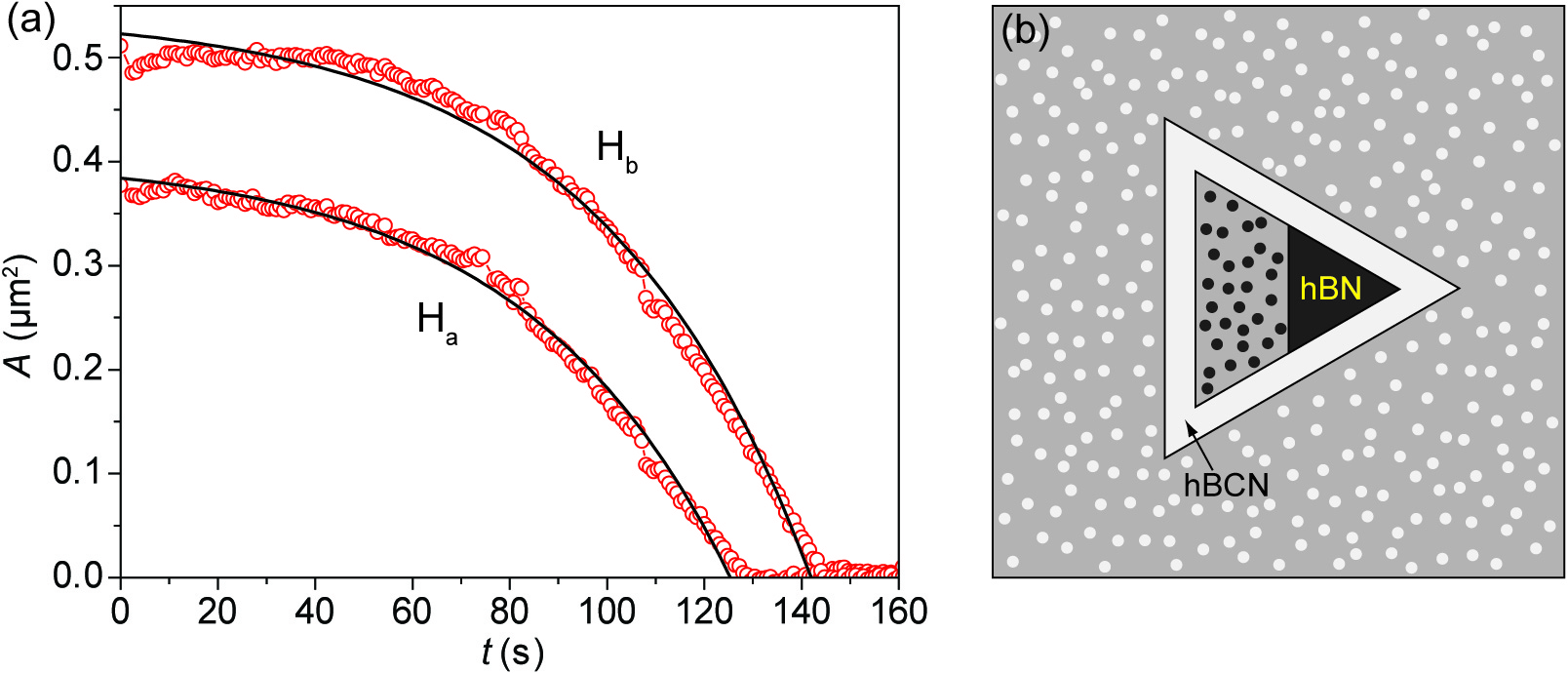}
\par\end{centering}
\caption{\label{fig5}(a) Red circles correspond to the area of the hBN cores
of the heterostructures H\protect\textsubscript{a} and H\protect\textsubscript{b}
shown in Figure \ref{fig4} as a function of annealing duration at
1370 K. Black lines are fits to the decay model given by Equations
\ref{eq1} and \ref{eq2}. (b) Illustration of a partially decomposed
heterostructure, with the in-equilibrium adsorbate species surrounding
hBCN region (bright dots) and the out-of-equilibrium species trapped
inside the forming microstructure (dark dots).}
\end{figure*}

Additional insight into disintegration of hBN cores during microstructuring
is obtained by monitoring their area ($A$). The resulting data is
plotted in Figure \ref{fig5}(a) as red circles for the two heterostructures
H\textsubscript{a} and H\textsubscript{b} shown in Figure \ref{fig4}.
The disintegration speed is relatively low in the beginning, but increases
with time until the hBN vanishes from the sample surface completely.
Apart from the different initial hBN core size, the data from H\textsubscript{a}
(triangular core) and H\textsubscript{b} (trapezoidal core) looks
similar, indicating that both hBN orientations present on Ir(111)
disintegrate in the same manner. We propose the following model to
describe the observed process.

Decomposition of hBN starts at the hBN-hBCN boundary, and further
on at the free hBN edge. Hence, to a first approximation and with
the aim of establishing a simplistic model, the decomposition rate
is proportional to the hBN area (the first term on the right hand
side of Equation \ref{eq1}) and becomes less significant as the time
progresses. However, this rate alone cannot explain the increasing
disintegration rate of hBN with time. That is why we introduce the
second component of the decomposition rate (the second term on the
right hand side of Equation \ref{eq1}) that relates to the local
concentration of the adsorbates as follows. If we take hBN-Gr heterostructure
with a partially decomposed hBN core, there will be a certain amount
of the adsorbed species (arising from decomposition) trapped in the
region bound by the remaining hBN and hBCN {[}see dark dots in Figure
\ref{fig5}(b){]}. As hBN decomposes further, the space available
to these species increases and their concentration decreases accordingly,
especially when taking into account desorption of B and N from the
surface and possible dissolution of B atoms in Ir at high temperatures
\cite{Zeiringer2015}. Due to the decrease of adsorbates concentration,
the system is constantly pushed into a non-equilibrium state, and
decomposition of hBN is further accelerated \cite{Rogge2015}. Therefore,
the second component of the hBN decomposition rate is proportional
to the area of already disintegrated hBN, and becomes relevant only
after an initial portion of hBN has already been removed from the
surface. The reduction of hBN area $A$ can then be written as 

\begin{equation}
\frac{dA}{dt}=-s_{1}A-s_{2}\left(A_{0}-A\right)\label{eq1}
\end{equation}

\noindent where $A_{0}$ is the area of hBN core prior to its decomposition
{[}$A_{0}=A\left(t=0\right)${]}. The solution to the above equation
is

\begin{equation}
A\left(t\right)=\frac{A_{0}}{s_{2}-s_{1}}\left[s_{2}-s_{1}e^{\left(s_{2}-s_{1}\right)t}\right].\label{eq2}
\end{equation}

Equation \ref{eq2} can be used to fit the data from Figure \ref{fig5}(a)
and calculate the values of $s_{1}$ and $s_{2}$. For the heterostructure
H\textsubscript{a}, we obtain $s_{1}=0.001$ s\textsuperscript{-1}
and $s_{2}=0.025$ s\textsuperscript{-1}, and for the heterostructure
H\textsubscript{b} we get $s_{1}=0.001$ s\textsuperscript{-1} and
$s_{2}=0.024$ s\textsuperscript{-1}. Therefore, the decomposition
rate linked to $s_{1}$ is an order of magnitude smaller than the
one linked to $s_{2}$, but it is essential in the initial stages
of the decomposition process. Overall, the values of $s_{1}$ and
$s_{2}$ for H\textsubscript{a} and H\textsubscript{b} are practically
identical, indicating generality of processes governing the hBN decomposition.

\begin{figure}
\begin{centering}
\includegraphics[width=8.2cm]{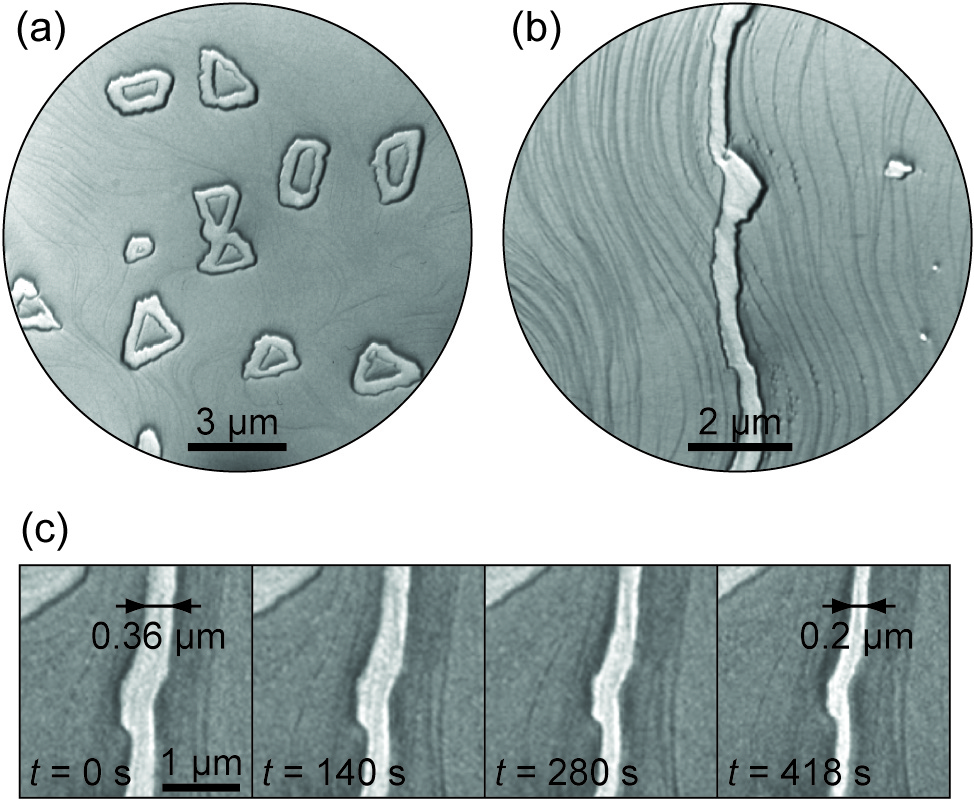}
\par\end{centering}
\caption{\label{fig6}Microstructures obtained by disintegration of hBN. (a)
Various hollow structures (synthesis at 1170 K, annealing at 1370
K) and (b) elongated stripe (synthesis at 1370 K, annealing at 1480
K). (c) An example of a stripe thinning process via oxygen etching.
$E=5$ eV.}
\end{figure}

A LEEM image of a larger area containing several microstructures is
shown in Figure \ref{fig6}(a), where the electron energy is chosen
to emphasize the shape of the microstructures. As before, triangular
and trapezoidal microstructures are visible. Also, an example of two
merged structures is found close to the center of the field of view.
If the heterostructure annealing step is performed at 1470 K instead
of 1370 K, partial degradation of Gr and hBCN can also take place.
Apparently, at this temperature sufficient non-equilibrium conditions
of the hBCN material and the surrounding adsorbed species {[}see bright
dots in Figure \ref{fig5}(b){]} can be reached, and significant number
of B, C and N atoms can detach from the hBNC crystal (atom detachment
rate becomes much larger than adsorbate attachment rate). As a result,
not only hollow but also concave Gr and hBCN microstructures can be
found on the Ir surface. If the size of the initial hBN core is sufficiently
large and its edges are straight enough, Gr and hBCN microstripes
can be fabricated. One such stripe is shown in Figure \ref{fig6}(b).
Similar as in the case of hollow microstructures, the thickness of
the stripe can be further modified by applying oxygen etching (thinning)
or CVD of ethylene (thickening), since both of these processes take
place at the edges of the existing Gr and hBCN structures. An example
of thinning is given in Figure \ref{fig6}(c) where an hBCN stripe
has been thinned down by etching in oxygen at 1170 K and a pressure
of $2.5\times10^{-8}$ mbar. The stripe is thinned from 0.36 μm to
0.2 μm during 418 s, yielding the average thinning speed of $\sim0.4$
nm/s.

\section{Discussion}

In sequential CVD of borazine and ethylene, C monomers and small C
clusters, originating from ethylene decomposition \cite{Loginova2009a,Tetlow2016},
are mixing with the adsorbed species that remain on the Ir surface
after hBN growth, as evident from Figure \ref{fig1}(e). These species
exist since islands of 2DMs must be in equilibrium with the surrounding
adatom or cluster gas (which serves as a feedstock during growth)
in order to be stable at a given temperature \cite{Loginova2008,Loginova2009a,Rogge2015}.
Therefore, possible candidates for the mixing with C are B atoms,
N atoms or various borazine fragments (either isolated or already
arranged in small hBN clusters). N atoms (including the ones forming
various molecules), B atoms and borazine molecules desorb from the
Ir(111) at temperatures relevant for our experiments, as was shown
in temperature programmed desorption (TPD) and x-ray photoemission
spectroscopy (XPS) experiments \cite{Cornish1990,Weststrate2010,Orlando2011}.
One additional source of mixing could be B atoms dissolved into the
subsurface region of Ir \cite{Usachov2012,Petrovic2017a}, but their
contribution must be minor since the alloy formation shows an obvious
dependence on the surface morphology of Ir (alloying in the step-down
direction is more pronounced), thus confirming the relevance of species
present on the surface.

This leaves dehydrogenated borazine fragments B\textsubscript{x}N\textsubscript{y}
and small hBN clusters as the most abundant species mixing with C
during heterostructure growth and formation of hBCN alloy. The C:(B,N)
ratio increases away from the hBN edge as the B\textsubscript{x}N\textsubscript{y}
fragments are incorporated into the growing heterostructure and become
depleted. Consequently, far enough from the hBN core, hBCN becomes
Gr, as is evident from the μ-LEED analysis presented in Figure \ref{fig3}.
Spatial variation of the C:(B,N) ratio also effects the electronic
structure of the hBCN alloy as visible in PEEM (see Supplementary
Figure S1), where variations in work function \cite{Xie2012} can
provide spatially-dependent photoemission yield. LEEM data {[}Figure
\ref{fig1}(a)-(d){]} clearly demonstrates more successful incorporation
of the B\textsubscript{x}N\textsubscript{y} fragments into the growing
hBCN alloy in the step-down direction of the Ir surface. This is analogous
to the growth of 2DMs on metal substrates that is often more effective
or even exclusive in the step-down direction \cite{Sutter2008,Loginova2009a,Sutter2011a}.
Accordingly, we speculate that the reason for anisotropic alloying
can be strong binding of hBN island edges to Ir step edges, which
hinders attachment of B\textsubscript{x}N\textsubscript{y} fragments
to the growing heterostructure in the step-up direction of Ir. This
is, however, not valid for carbon species that are incorporated into
the heterostructure also in the step-up direction, which is a clear
indication of different growth mechanisms of hBCN alloy and graphene
regions. Also, pronounced anisotropy in diffusion of the B\textsubscript{x}N\textsubscript{y}
fragments and small hBN clusters (differences in step-up, step-down
and along-the-terrace diffusion) might cause local inhomogeneities
in material supply, and contribute to the alloy non-uniformity and
appearance of distinct graphene regions within the hBCN alloy {[}see
white arrow in Figure \ref{fig1}(d){]}. Overall, isotropic alloying
could be achieved if more advanced methods for substrate preparation
would be employed (that would result in low density of surface steps),
which leaves room for further advancement in the material quality.

Moiré diffraction spots are always visible in μ-LEED analysis (see
Figure \ref{fig3}), proving that besides pure hBN and Gr regions,
hBCN alloy also exhibits a hexagonal lattice. We do not observe simultaneous
presence of the hBN and Gr moiré spots, which indicates that indeed
the transition region from hBN to Gr is a locally homogeneous alloy.
However, it is possible that separated hBN and Gr nano-domains, too
small to provide a clear diffraction pattern, exist within hBCN, since
calculations predict that phase segregation is energetically favorable
\cite{Lam2011,DaRochaMartins2011,Manna2011,Xie2012}. It can be inferred
from Figure \ref{fig3}(a) that the lattice constant of hBCN alloy
changes from almost 2.48 Å \cite{FarwickZumHagen2016} {[}lattice
constant of hBN on Ir(111){]} to 2.45 Å \cite{NDiaye2008} {[}lattice
constant of Gr on Ir(111){]} as a function of the distance from the
hBN edge. In a recent μ-LEED study, Camilli et al. find that the hBCN
alloy has a lattice constant of 2.57 Å when analyzing more complex
samples consisting of Gr nanodots embedded in an hBCN matrix \cite{Camilli2017}.
This suggests that different synthesis procedures can provide hBCN
alloys of different stoichiometries and lattice constants. Modification
of lattice constant in our experiments is accompanied by the in-plane
rotations across the heterostructure, which we ascribe to lattice-matching
process arising from the continuous variation of the hBCN lattice
constant. Similar lattice miss-orientations were found in Gr-hBN heterostructure
synthesis on Cu foils by atmospheric pressure CVD \cite{Han2013}.

During the decomposition of hBN and microstructuring, N atoms desorb
from the surface, most probably forming various molecules since N
is volatile. B atoms, especially at temperatures above $\approx1220$
K, can diffuse into the subsurface regions of the Ir, they segregate
to the surface after the sample is cooled, and are then easily detected
with LEEM \cite{Petrovic2017a}. But since no significant segregation
of B has been observed during cooldown, we conclude that a large portion
of B atoms also desorbs from the Ir surface at high annealing temperatures.
In addition, if the temperature increase is rapid as in our experiments,
the time available to B atoms to dissolve into Ir, before they have
enough energy to desorb from the surface, is limited. This further
supports our model of hBN decomposition, which depends on the concentration
of adsorbates on the Ir surface. The hBN-Gr heterostructures and the
corresponding microstructures are isolated from each other and are
not mutually interconnected or overlapped, and are therefore suitable
for further integration into more complex systems. It is worth noting
that a few other studies also reported on the fabrication of hollow
2D structures on Pt and Cu foils by growth-etching procedures in atmospheric-pressure
tube furnaces, namely hexagonal Gr rings \cite{Ma2013,Liu2016a} and
triangular hBN structures \cite{Sharma2015}.

\section{Conclusion}

In sequential CVD growth of lateral hBN-Gr heterostructures on Ir(111),
a 2D hexagonal alloy consisting of B, C, and N (hBCN) is formed at
the interface between hBN and Gr. Alloying occurs because in the first
synthesis step epitaxial hBN islands are formed that are in equilibrium
with the surrounding borazine fragments, and these fragments mix with
C species that originate from ethylene decomposition taking place
in the second synthesis step. The C:(B,N) ratio within the alloy,
as well as the width of the alloy region, can be modified by altering
the temperature and the ethylene pressure during synthesis. In this
way, it is possible to tune the electronic properties (i.e., the band
gap and the work function) and spatial arrangement of the alloyed
material, which provides means for customization that is crucial in
many applications. In addition, introduction of oxygen into the synthesis
procedure enables production of sharp hBN-Gr interfaces without the
hBCN region. Due to the decomposition of hBN at elevated temperatures,
it is possible to fabricate various hBCN and Gr microstructures upon
annealing of hBN-Gr heterostructures. This kind of hBN-templating,
in combination with well-known transfer techniques, effectively provides
foundations of a new method for production of micro- and nano-sized
conducting and semiconducting elements that can be integrated in more
complex devices (e.g. transistors, solar cells, light emitters or
resonators). The advantage of this method over more established ones,
e.g., lithography, is that it does not require undesirable ex situ
manipulation of the sensitive atomic layers, structures in the 100
nm range can be produced with ease, and hence we envision its further
development as an alternative route for technological implementation
of 2DMs.

\section{Acknowledgments}

M.P. would like to thank the Alexander von Humboldt Foundation for
financial support.

\section{References}

\bibliographystyle{elsarticle-num}
\bibliography{references}

\end{document}